\RequirePackage{fix-cm}
\documentclass[smallextended]{svjour3}       % onecolumn (second format)
\smartqed  % flush right qed marks, e.g. at end of proof
\usepackage{graphicx}
\usepackage{multicol}
\usepackage{color}
\usepackage{natbib}
\usepackage{amsmath, amssymb}
\usepackage{booktabs}
\DeclareMathOperator*{\argmax}{arg\,max}

\usepackage{comment}

\usepackage[]{algorithm2e}
\SetKw{KwBy}{by}
\SetKwFor{For}{for (}{) do}{end}

\newcommand{\dd}{\text{d}}

\begin{document}

\title{Learning Social Networks from Text Data using Covariate Information
%\thanks{}
}
% Grants or other notes about the article that should go on the front
% page should be placed within the \thanks{} command in the title
% (and the %-sign in front of \thanks{} should be deleted)
%
% General acknowledgments should be placed at the end of the article.


\author{Xiaoyi Yang        \and
        Nynke M.D.\ Niezink      \and
        Rebecca Nugent %etc.
}

\authorrunning{ } % if too long for running head

\institute{X.\ Yang \at
              Department of Statistics \& Data Science \\
              Carnegie Mellon University \\
              Pittsburgh, PA 15206, USA \\
              Tel.: +1608-622-4955 \\
              \email{xiaoyiy@andrew.cmu.edu}           %  \\
           \and
           N.M.D.\ Niezink \at
              Department of Statistics \& Data Science \\
              Carnegie Mellon University \\
              Pittsburgh, PA 15206, USA
            \and
            R.\ Nugent \at
               Department of Statistics \& Data Science \\
               Carnegie Mellon University \\
               Pittsburgh 15206, PA, USA
}

\date{Received: date / Accepted: date}

\maketitle

\newpage

\begin{abstract}

Describing and characterizing the impact of historical figures can be challenging, but unraveling their social structures perhaps even more so. Historical social network analysis methods can help and may also illuminate people who have been overlooked by historians, but turn out to be influential social connection points. Text data, such as biographies, can be a useful source of information about the structure of historical social networks but can also introduce challenges in identifying links. The Local Poisson Graphical Lasso model leverages the number of co-mentions in the text to measure relationships between people, and uses a conditional independence structure to model a social network. This structure will reduce the tendency to overstate the relationship between ``friends of friends'', but given the historical high frequency of common names, without additional distinguishing information, we can still introduce incorrect links.  

%Conditional independence structure like Poisson Graphical Model, which makes use name mention counts in the text can be useful tools to avoid false positive links due to the co-mentions but given historical tendency of frequently used or common names, without additional distinguishing information, we may introduce incorrect connections. 

In this work, we extend the Local Poisson Graphical Lasso model with a (multiple) penalty structure that incorporates covariates giving increased link probabilities to people with shared covariate information. We propose both greedy and Bayesian approaches to estimate the penalty parameters.  We present results on data simulated with characteristics of historical networks and show that this type of penalty structure can improve network recovery as measured by precision and recall. We also illustrate the approach on biographical data of individuals who lived in early modern Britain, targeting the period from 1500 to 1575.

%We will show how these covariates affect the statistical model's performance using simulations, discuss how it helps to better identify links for the people with common names and those who are traditionally underrepresented in the biography text data. 

\keywords{Local Poisson Graphical Lasso model \and Social networks \and Text data \and L1 penalty factor \and Bayesian penalty estimation}
\vspace{0.2cm}
\noindent\textbf{Declaration}
\\
Funding: Not applicable
\\
Conflicts of interest/Competing interests: Not applicable
\\
Availability of data and materials: The data used in Section \ref{s:example} is part of the Six Degrees of Francis Bacon project (SDFB). Contact \cite{warren2016six} for the data. 
\\
Code availability: All code corresponding to the paper is be available upon request.

\end{abstract}

\newpage
\section{Introduction}
\label{intro}
The structure between individuals reflecting their social relationships can be represented using networks. In modern society, learning social networks is relatively simple, since accessing information about or from each individual can be done through multiple sources. For example, to learn someone's personal social network, we could either survey them or use information from social media platforms. In an academic setting, co-authorship information from curriculum vitae, journals, or online archives can provide information about collaborations. However, things become more difficult if we are interested in learning a historical social network, for which the potential sources of information are limited.  Building networks based on text data may identify less well-known figures in potential social connection positions of influence. In this paper, we will present methodology to unravel (historical) social networks, based on text data. 

Previous work on estimating social networks from text data spans different humanities and social science applications. For example,  \cite{marsden1990network}  and \cite{usdiken1995organizational} use structured surveys and citations respectively to estimate sociological and collaboration networks. \cite{almquist2019using} uncover the underlying network structure of radical activist groups with British radical environmentalist texts from 1992 to 2003. Their work primarily concentrates on the application of topic models to analyze the text, and they infer a network using text co-occurrence counts. The China Biographical Database Project\footnote{See {\texttt{https://projects.iq.harvard.edu/cbdb/home}}.} is also a great example of how networks can be extracted from historical documents. This labor-intensive project involved manually listing all the possible expressions of human relations (e.g. ``A is friends with B'') and then searching the text using pattern matching to extract relational links. Social networks can also be estimated from fiction texts: \cite{bonato2016mining} extract and analyze the social network from three best-selling novels, defining a link between two characters if their names co-appear within 15 words.  %This approach still requires parsing for names and aliases prior to searching for co-appearances.

\subsection{Six Degrees of Francis Bacon}

Six Degrees of Francis Bacon \footnote{See {\texttt{http://www.sixdegreesoffrancisbacon.com/}}} (SDFB) is a recent historical network project, focusing on estimating the social network in early modern Britain during 1500--1700 \citep{warren2016six}. As their text source, SDFB uses biographies from the \textit{Oxford Dictionary of National Biography}\footnote{See { \texttt{https://www.oxforddnb.com/}}} (ODNB) and represents possible relationships between people by the number of times the name of one person occurs in a section of the other person's biography, under the assumption that if two people knew each other as more than just acquaintances and/or were colleagues, they are more likely to show up in each other's biographies.  \cite{warren2016six} uses a Local Poisson Graphical Lasso model \citep{allen2012log} to estimate the social network using these count data, arguing that a conditional independence structure should be considered when constructing these types of historical social networks. This will help distinguish whether two people truly know each other or just happen to be co-mentioned in a document. 

There is no doubt that SDFB contributed a rich resource to support humanities research on early modern Britain. However, given this auspicious start, there is room for improvement. For example, in their validation of precision and recall among 12 non-random people, \cite{warren2016six} find that the SDFB approach tends to have high precision but relatively low recall. One possible reason for this behavior is that the model only makes use of the co-mention counts but ignores other information available in the text such as individual characteristics (e.g., occupation, social group). According to homophily theory, similar individuals are more likely to connect to each other than the dissimilar ones \citep{mcpherson2001birds}. For example, several studies have shown that people who share similar age, education level \citep{kossinets2009origins}, occupation \citep{calvo2004effects}, gender and economic status \citep{mcpherson1982women} are more likely to be connected. Indeed, when looking at SDFB's reconstructed network, we can also observe that linked people tend to share some common characteristics. Including information such as last name or profession/social group in the model may help us increase link accuracy.

%so including the information like common last name and collective group identities should be able to help us pick more accurate links than what SDFB has done.

%Therefore, we need a model to learn networks from text data that incorporates more information in the text than just the co-mentions. The addition information, for example in the ODNB, can be a person's last name, birth and death year, occupation and social group memberships. We consider each piece of the information as a covariate and in a network model, we assume people who share more common covariates should be more likely to know each other. 

When adding additional covariates to our network model, there are several factors we need to consider. First, the model should be flexible enough to include the large number and variety of covariates available in historical text data.  Second, we need to include possible aggregation of covariate effects. For example, sharing both last name and occupation likely has a different impact on a possible relationship than just sharing last name. % Third, covariate comparisons should not be restricted to just binary, e.g. match vs non-match.  For example, there may be typographical errors or name variations.  %Similarly, sharing a rare last name is likely a stronger signal than sharing a common last name.

%Third, the comparison of covariates are not necessary binary. For instance, the people who share a rare last name should be more likely to be connected compare to the people who share a common last name. 

%First, there can be a large number of covariates exist in the text, so we want the model can easily contain any number and combination of covariates and estimate efficiently. 

The idea to incorporating additional information into the Lasso regression model is not new. \cite{yuan2006model} proposed group Lasso to add penalties to groups rather than individuals. \cite{li2015multivariate} extended this method to a multivariate sparse group Lasso to incorporate arbitrary and group structures in the data. Their model provided a unique penalty for each node but also include a penalty for each group where the groups could overlap and even be nested. \cite{zou2006adaptive} proposed the adaptive LASSO which uses initial coefficient estimates without regularization to inform starting penalty weights. However, these papers did not include approaches for incorporating additional information into the penalties outside of group structure. 

On the other hand, \cite{boulesteix2017ipf} proposed IPF-Lasso which assigns different penalty factors to all independent variables in their model that are a function of external information, and use cross validation to select penalty parameters based on model performance. In a similar vein, \cite{zeng2020incorporating} use the Bayesian interpretation of the penalized regression, re-formulating Lasso regression as a Bayesian model. % Each coefficient of the original Lasso regression is now modeled with a Laplace distribution. The Laplace distribution has a mean 0 and the scale parameter is proportional to the individual penalty for each feature in the regression. Such individual penalty is modeled as a log-linear function of the covariates. 
However, these approaches have not been implemented in the Poisson case, which involves different estimation challenges.

%In additional to above, there are also several approaches in historical and social science field to identify social network links in the text data. The text data can be in various forms. 

In this paper, we explore and extend both \citeauthor{boulesteix2017ipf} and \citeauthor{zeng2020incorporating}'s approaches to the Local Poisson Graphical Lasso model in the context of the SDFB project. We will show (1) how to implement the node-wise covariates and information into the network model's penalty factors, (2) two potential methods for estimating penalty factors for potentially a large number of covariates, and (3) how the inclusion of additional information into penalty estimation can significantly improve precision and recall.

\section{Model}
We consider text data consisting of $n$ documents with approximate similar length, based on which we aim to learn the social network among $p$ individuals. This type of information can be obtained by manually coding documents or using natural language processing techniques. We assume that if two people are mentioned in the same document, then they are more likely to know (or have known) each other. More specifically, if two people's counts of name mentions in a document, conditional on all other people's mentions, are positive correlated, then this is an indication that they may know each other.  Note that even if two individuals are co-mentioned in a document, this may not imply that they directly know each other:  they could, for example, have a common acquaintance. For this reason, conditional independence structures are a natural tool for estimating social networks from text data. 

Let $Y \in \mathbb{N}_0^{n\times p}$ be the document by person matrix where $Y_{ij}$ indicates how many times person $j$ is mentioned in document $i$.  Each row represents the counts of all name mentions in a document; each column represents the name mentions for one person across all documents.  We denote an observed document by person matrix by $y$, while $Y$ denotes the random matrix.

%Here, we extend the Local Poisson Graphical Lasso model, which was also used in the SDFB study .

%We assume that if two individuals knew each other, they are more likely to show up in the same document or the same part of the document. 

The Local Poisson Graphical Lasso model \citep{allen2012log,warren2016six} is a variant of the Poisson graphical model that enforces sparsity and is estimated locally. The Poisson graphical model assumes that each name count in a document, conditional on all the other name counts in that document, is Poisson distributed. For document $i$, let $Y_{i, \neq j}$ denote the vector of name counts for all individuals other than $j$. The Poisson graphical model can be expressed as
\begin{equation} \label{eq:model}
    Y_{ij}\,|\,Y_{i,\neq j} = y_{i,\neq j}, \theta, \Theta \sim \text{Poisson}(e^{\theta_j + \sum_{k \neq j}  y_{ik}\Theta_{kj}}),
\end{equation}
with $\theta \in \mathbb{R}^{p}$ and $\Theta \in \mathbb{R}^{p\times p}$ ($\Theta_{ii} = 0$ for all $i$) and where node parameter $\theta_j$ serves as the intercept and the edge parameters $\Theta_{jk}$ indicate the relations between individuals $j$ and $k$. If $\Theta_{jk} > 0$, then if individual $k$ is mentioned often in a document, $j$ is likely to be mentioned often as well, which is suggestive of a social tie between $j$ and $k$. The opposite is true when $\Theta_{jk}$ is negative. We thus consider $\Theta_{jk} > 0$ as an indication of a social tie between individuals $j$ and $k$.

\subsection{Penalty factors}
\label{pf}

The Poisson graphical model leverages the name count data in text to learn social network information. However, generally, the body of text in which names are embedded is rich with other information that could be indicative of social ties.  
For example, it is useful to consider demographic information available from the text when reconstructing social networks from text data. People with common characteristics are generally more likely to be connected than those who are not alike. When learning historical networks based on text data, it might be relevant to know, e.g., whether individuals were part of the same family or social group/club, worked for the same company, and whether they lived geographically close to one another.

Here we extend the Local Poisson Graphical Lasso model with a multiplicative factor for the penalty term that depends on individual covariate information inferred from the text. For person $j$, we define the covariate matrix $Z^j \in \{0,1\}^{p \times m}$, with $m$ covariates, by
\begin{equation}
    Z^j_{kh} = \begin{cases} 
        1 & \text{if person $j$, person $k$ have an equal value for covariate $h$}, \\
        0 & \text{otherwise.}
    \end{cases}
\end{equation}

We here consider binary-valued matrices $Z^j$, but the approach proposed in this paper is also applicable to real-valued covariates. Examples of these include last name similarity and last name or social group commonality scores.

For each covariate we include a different penalty factor. Thus, for each person $j$, the penalized Lasso estimators are given by
\begin{equation} \label{eq:penalizedestimator}
    \begin{aligned}
    \hat \Theta_j =  \argmax_{\Theta_j} &\sum_{i=1}^n \left[y_{ij}(y_{i, \neq j}\Theta_{\neq j, j}) - e^{y_{i, \neq j}\Theta_{\neq j, j}}\right] \\ &- \sum_{k \neq j}  |\rho_{k j}\Theta_{k j}| \quad \text{with }
    \log(\rho_{\neq j, j}) =  Z^{j*} \alpha,
    \end{aligned}
\end{equation}	
where $Z^{j*}$ is the matrix $Z^j$ with the $j$th row taken out and prefixed by an all-one column vector and $\alpha \in \mathbb{R}^{m+1}$ denotes the penalty factor. The first element of $\alpha$ is $\alpha_0$, an intercept controlling the overall shrinkage. If two individuals $k$ and $j$  share a common value on a covariate $h$, the penalty for parameter $\Theta_{jk}$, indicating the link between them, is $e^{\alpha_h}$ times the overall penalty. Therefore, if having covariate $h$ in common makes two people more likely to be connected, then $\alpha_h$ will be negative. Otherwise, it will be positive.  

Birth and death date are covariates that deserve special treatment in this framework, since if two individuals were not alive at the same time, they could not have had a social connection. To address this, \cite{warren2016six} removed the links between people who were not alive at the same time post-network estimation. Given our penalty factor structure, we can instead include birth and death year information directly into the model. We set the penalty factor for the lifespan overlap covariate to be infinity and so do not link people with non-overlapping birth and death years.

%By setting the penalty factor for the covariate that has value 1 if two individuals were not alive at the same time to infinity, the link between them will receive an infinite penalty. 

\section{Estimation} \label{s:estimation}

For each person $j$, we fit a Poisson regression model including an L1 penalty to enforce sparsity. We estimate model parameters via penalized maximum likelihood using cyclical coordinate descent, as implemented in the R package \texttt{glmnet} \citep{friedman2010regularization}. This method consecutively optimizes the objective function given as part of expression \eqref{eq:penalizedestimator} over each parameter while keeping the others fixed, and cycles until convergence. 

During estimation, we force the coefficients to be non-negative, as opposed to \cite{warren2016six} who estimated a local Poisson graphical model and ignored the negative coefficients after estimation. After estimating the edge parameter vectors for persons $j$ and $k$, we say that there is a social tie between $j$ and $k$ when at least one of $\hat\Theta_{jk}$ and $\hat\Theta_{kj}$ is positive. 

Estimating the value of penalty vector $\alpha$ is essential for determining the edge parameters. In the following two sections, we discuss two approaches to estimate the penalty vector: using greedy search (Section \ref{s:greedy}) and using the reformulation of Lasso regression in the Bayesian framework (Section \ref{s:bayes}). 

\subsection{Greedy search} \label{s:greedy}

One way to estimate the penalty factor $\alpha$ is by defining a grid of penalty parameter values and evaluating the corresponding models, selecting the values that minimize the prediction error \citep{boulesteix2017ipf}. However, this approach is generally computationally feasible only when the number of covariates is small (say, no more than four). Greedily searching the parameter space allows for inclusion of more covariates. Our proposed greedy algorithm for $\alpha$ is described in Algorithm \ref{Greedy}. 
Starting with all $\alpha_h$ = 0, i.e. no penalty adjustment, the algorithm first iterates over all covariates in random order.  For each covariate $h$ and a gridded range of pre-specified $\alpha_h$ values, we use cross-validation to choose the baseline $\hat{\alpha}_0$ (holding all other $\alpha_h$ penalty parameters fixed) and calculate the corresponding MSE. We then choose the $\hat{\alpha}_h$ corresponds to the lowest MSE. After randomly iterating through all covariates, the algorithm repeatedly randomly iterates through all covariates again, looking for possible updated $\hat{\alpha}_h$ values, stopping when no further $\alpha_h$ tuning leads to a decrease in MSE.

%in case any of the $\alpha_h$'s need to be updated, only stopping when no $\alpha_h$ tuning leads to a decrease in MSE. 

%we first calculate the mean squared error (MSE) of each $\alpha_h$ from a gridded range of prespecified values, while keeping all other penalty parameters fixed, through cross-validated of the overall penalty $\alpha_0$. We select the $\hat{\alpha}_h$ that minimizes the mean squared error (MSE). 

\begin{algorithm}[!ht]
\DontPrintSemicolon
\SetKwInOut{Input}{Input}
\SetKwInOut{Output}{Output}
\Input{document-by-person matrix $Y$, covariate matrices $Z^{j*}$ for $j=1,\ldots,p$,\; 
\phantom{\textbf{Output: }}function MSE$(\alpha$) evaluating the MSE for the model with penalty $\alpha$,\;
\phantom{\textbf{Output: }}search range $[a_h, b_h]$ for $h = 1,\ldots,m$, grid size $d$\;
%phantom{\textbf{Input: }} a vector s to record a random permutation of 1 to $m$
}
\Output{$\hat \alpha = [\hat\alpha_1,\ldots, \hat\alpha_m]$}
\BlankLine

 \textbf{Initialization:} $\hat \alpha \leftarrow [0,0,...0]$, repeat $\leftarrow $ true,\;
 \phantom{\textbf{Initialization: }}$MSE_{old} \leftarrow MSE(\hat \alpha)$,
 $MSE_{new} \leftarrow \infty$\;
\BlankLine
 \While{{\upshape repeat}}{
   repeat $\leftarrow$ false \;
   order $\leftarrow$ a random permutation of 1 to $m$ \;
  \For{$ s \gets 1 $ \KwTo $ m $ \KwBy $1$}{
    h $\leftarrow$ order[s]\;
  \For{$\hat  \alpha_h^* \gets a_h $ \KwTo $ b_h $ \KwBy $d$}{
    $\hat \alpha^* \leftarrow [\hat \alpha_1,..., \hat  \alpha_h^*,..., \hat \alpha_m]$ \;
    $MSE_{new} \leftarrow MSE(\hat \alpha^*)$\;
    \If{$MSE_{new} < MSE_{old}$}{
    $MSE_{old} \leftarrow MSE_{new}$\;
    $\hat \alpha \leftarrow \hat \alpha^* $\;
    repeat $\leftarrow$ true
    }
    }
    }
 }
\BlankLine
\BlankLine

\caption{Greedy algorithm to estimate $\hat \alpha$}
 \label{Greedy}
\end{algorithm}

To use Algorithm \ref{Greedy}, we need to specify the search range for $\alpha_h$ and the step size $d$. We recommend starting with a search range for $\alpha_h$ such as $[-1.2, 0.5]$ (values outside that range have diminishing impact on the multiplicative factor value) and a relatively large step size $d$ (e.g., 0.1). The search range can be enlarged if the margins are hit during the initial estimation. Decreasing the step size $d$ can of course lead to a more fine-grained solution but will depend on any present computational constraints. We could also choose the search range $[a_n, b_n]$ using prior information, such as which covariates are expected to be influential and approximately how they might affect the chance of two individuals to be connected. For example, if we know a covariate $h$ is likely associated with an increased chance of a link, we could initially limit the search range of $\alpha_h$ to the negative numbers. This type of search range adaptation can also decrease the number of algorithm iterations and computational time.

%Using such information helps to narrow the search range thus decreasing the number of runs within the algorithm. 

%Enlarge the search range, in case the margins of the search range are hit during the initial estimation. By subsequently decreasing the step size $d$, one may get a more accurate estimation.

\subsection{Bayesian estimation}  \label{s:bayes}

Lasso estimates can equivalently be derived as the Bayesian posterior modes under independent Laplace priors for the parameters to which shrinkage is applied \citep{tibshirani1996regression}. Therefore, we can use the Bayesian framework to estimate the penalty parameters $\alpha$.
To this end, we complement model \eqref{eq:model} with a Laplace prior for the edge parameters: for $k\neq j$,

\begin{equation} \label{eq:laplaceprior}
        \Theta_{kj} \sim \text{Laplace} \left(0, b_{kj}\right), \qquad  b_{kj} \propto \rho_{kj}, \qquad \log(\rho_{\neq j,j}) =  Z^{j*} \alpha.
\end{equation}
Notice that $\alpha$ only influences the penalties on the edges and not the node parameters $\theta_j$. We will specify the exact form of $b_{kj}$ later in this section.
We here extend work by \cite{zeng2020incorporating} 
on incorporating covariate-dependent penalty factors in the Lasso term in linear regression and linear discriminant analysis (LDA) models to the Local Poisson Graphical Lasso model.

We use an empirical Bayesian approach to estimate the penalty parameters $\alpha$. First, for each person $j$, we approximate the marginal log-likelihood of $\alpha$,  denoted by $l_j(\alpha)$, marginalizing over the coefficients $\Theta_{\neq j, j}$. The estimate of $\alpha$ is given by
\begin{equation}\label{eq:marginallik}
\hat \alpha = \argmax_{\alpha} \sum_{j=1}^p l_j(\alpha). 
\end{equation}
Note that we maximize the sum of the marginal distributions, because we need a global penalty factor over all people instead of for one specific person $j$. Since the $l_j(\alpha)$ are not convex, we use a Majorization Minimization procedure \citep{zeng2020incorporating} to estimate $\hat \alpha$. We then use the $\hat \alpha$ as input for the penalized maximum likelihood estimation of the model, as summarized at the start of Section \ref{s:estimation}.

Since the Poisson regression likelihood and the Laplace prior are not conjugate pairs, there is no closed form expression for the marginal likelihood of $\alpha$. We here present a general outline of how we approximated $l_j(\alpha)$, approximating both the Poisson regression likelihood and the Laplace prior -- see Appendix A for the full derivation.
First, we apply the log-gamma transformation to approximate the Poisson regression likelihood by a multivariate Gaussian distribution \citep{chan2009bayesian}. In order to avoid $\log(0)$ in our derivation, we add 1 to all the observed outcomes $y_{ij}$, that is, define $y_{ij}^* = y_{ij} + 1$. 
%Such modification should only affects the node parameter, which will not affect the value of $\alpha$. Then the Poisson likelihood can be written as
%\begin{equation}
%    \begin{aligned}
%    \prod_{i=1}^n p(Y_{ij}\mid Y_{i, \neq j} = y_{i, \neq j}, \Theta_{\neqj, j}) &= \prod_{i=1}^n \frac{1}{y_{ij} !}e^{\lambda(y_{i, \neq j}) y_{ij}} \exp^{-e^{\lambda(y_{i, \neq j}) }} \\
%    &\approx \prod_{i=1}^n \text{G}(\lambda(y_{i, \neq j}) \mid \log(y_{ij} + 1), \sigma^2_j) \\
    %&\approx \frac{|\sigma_j I_n| ^{-1/2}}{(2\pi)^{\frac{N}{2}}} e^{-\frac{1}{2}|| \lambda(Y_{ \neq j})- \log(Y_{j} + 1)||^2_{\sigma_j I_n}}
%    \end{aligned}
%\end{equation}
%where $\lambda(y_{i, \neq j}) = \theta_j + \sum_{k \neq j} y_{ik} \Theta_{kj}$, $G(x|\mu, \Sigma) = (2\pi)^{-d/2} |\Sigma|^{-1/2} \exp(\frac{1}{2}|| x - \mu||^2_\Sigma$ is the equation of a multivariate Gaussian distribution, $||x||^2_\Sigma = x^T \Sigma^{-1} x$, $\sigma^2_j = \frac{1}{n} \sum_{i=1}^n \frac{1}{y_{ij} + 1}$.

Second, we assume $\Theta_{kj}$ follows the  Laplace prior $\Theta_{kj} \sim \text{Laplace} (0, \frac{\rho_{kj}}{2\sigma_j^2})$, where $\hat\sigma_j^2 = \sum_{i=1}^n \frac{1}{y_{ij}^*}$ is the estimated variance in Gaussian distribution approximating the Poisson likelihood. We approximate this prior by a normal distribution with the same variance \citep{zeng2020incorporating}, yielding
\begin{equation}
\Theta_{\neq j, j} \sim \mathcal{N}(0, V^j)
\end{equation}
where $V^j \in \mathbb{R}^{(p-1)\times (p-1)}$ is a diagonal matrix with 
$V^j_{kk} = 2\sigma_j^2\exp^{-2Z^{j*}_k\alpha} $, 
in which $Z^{j*}_k$ is the $k$th row of the covariate matrix $Z^{j*}$.

Combining the two, we can approximate the log-likelihood of $\alpha$ for person $j$ and find
\begin{equation}
    -l_j(\alpha) \propto \log|C_{\alpha}| + \log(y_{j}^*)^\top C_{\alpha}^{-1} \log(y_{j}^*)
    \label{objective}
\end{equation}
where $C_\alpha = \sigma_j I^2 + y_{\neq j} V^{j}y_{\neq j}^\top$,  $y_{\neq j}$ denotes data matrix $y$ excluding the $j$th column, and $\log(\cdot)$ is applied element-wise to  $y_{j}^* = (y_{1j}^*, \ldots, y_{nj}^*)^\top$. Integrating this in expression \eqref{eq:marginallik}, we can estimate the penalty factors.
    
\section{Simulation study}

In order to compare the methods quantitatively, we created a small community (network) as the ground truth for the simulation. The community is similar by design to the SDFB social network, and we assume to have similar available demographic information. When creating the network, we considered three covariates: last name, group membership and birth/death year overlap. Note that we assume that if the lifespan of two people does not overlap (i.e., it is impossible that they physically met each other), then they should not be linked, regardless of all other factors. Below is a description of the general network design:
\begin{enumerate}
    \item We generate 50 families in the community with 30 different last names (i.e., people with the same last name can be from different families). 
    \item For each family, we randomly generate 5 to 12 people, each with a birth and death year between 1500 and 1600 and a life length varying from 5 to 70. 
    \item Within a family, among those people whose lifespan overlaps, 50\% know each other.
    \item There are three social groups, A, B and C. Each person is randomly assigned to one of the groups with probability 0.5, 0.25 and 0.25, respectively.
    \item Among those people whose lifespan overlaps, we additionally create 100, 100 and 50 links within group A, B and C, respectively. 
    \item At the end, we add 300 random links to the whole community. 
\end{enumerate}

\begin{figure}[!ht]
    \centering
    \includegraphics[width=0.5\textwidth]{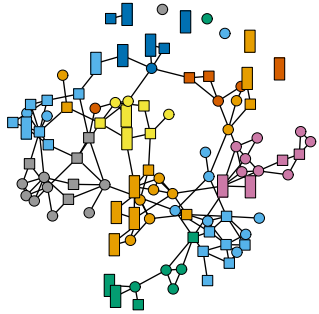}
    \caption{10-family sub-community (100 people, 158 links), Last names are represented by colors and social groups by shapes. The network shows clear family structure with some social group structure and additional random links.}
    \label{fig:simu_network}
\end{figure}

This design yields 464 people and 1164 links. Figure \ref{fig:simu_network} illustrates a subset of the community with ten families, 100 people, and 158 links. Since our network design yields higher link density within a family/group as compared to the network average, we anticipate all $\alpha$ should be negative, thus lead to a smaller penalty if two people share the same last name or are in the same group. 
From Figure \ref{fig:simu_network}, we can see the expected network structure associated with families; we expect the absolute value of the penalty factor for last name to be large. The social group covariates might also be significant but should have a relatively smaller effect than last name. Since group B is the densest one, we expect that its penalty factor may be larger compared to the other two groups. If two people's birth/death year do not overlap, we automatically set the penalty to be infinity, leaving us only four covariates to consider: last name, group A, group B and group C. 
As described in section \ref{pf}, we build a series of covariate matrices $Z^j$ using the criterion of whether person $j$ and person $k$ have the same exact covariate $h$ value.

Using a simulation framework adopted from \cite{allen2012log}, we then generate 10 different document-by-person matrices, each with 2000 documents (i.e. of similar size as SDFB). For each matrix, we run the greedy algorithm and Bayesian approach to estimate the best penalty factor $\alpha$. Since the matrices are all generated from the same network, we assume the $\hat{\alpha}$ should be relatively similar and reflective of the designed network structure (relatively larger covariates for last name and group B). We then compare the network estimated with no penalty adjustment to the networks estimated with penalty adjustment (greedy, Bayesian) by calculating their precision and recall.  Our expectation is that the penalty adjustment allowing for incorporating covariate information into the network estimation will be associated with an improved average of precision and recall.   We also examine the predicted 10-family sub-communities in an attempt to characterize how the penalty factors impact the false positives and false negatives.

%Then, we estimate the network without penalty parameter, with the penalty parameter from greedy algorithm, and with the penalty parameter from Bayesian approach respectively, and calculate the precision and recall for each case. 
%We are assuming that with the penalty parameter, the average of precision and recall should be improved.
%Moreover, we will also present two predicted sub-network, to show including the penalty factor does help us to achieve a better representation of the predicted network, and try to understand how the penalty factors affect the false positive and false negative. 

\begin{figure}[!ht]
    \centering
    \includegraphics[width=1\textwidth]{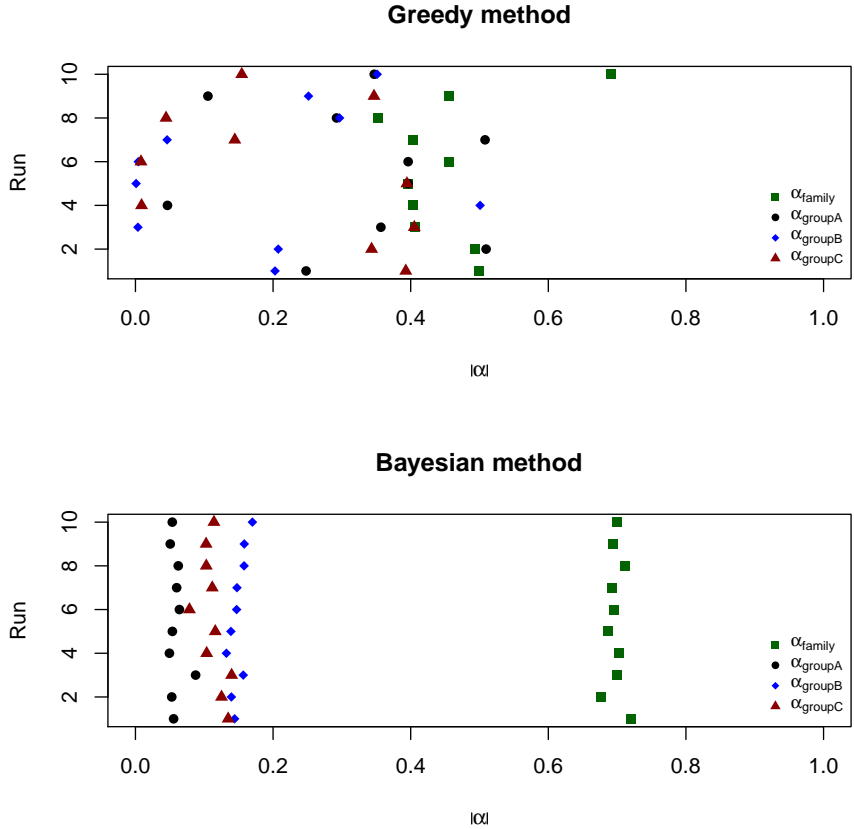}
    \caption{The distribution of $|\hat{\alpha}_h|$ estimated by the greedy and Bayesian approaches over ten runs. Both approaches generally pick the last name as the most important covariate. The Bayesian approach tends to give more consistent values while the greedy approach estimates have a larger variance. }
    \label{fig:alpha}
\end{figure}

Figure \ref{fig:alpha} shows the distributions of $|\hat{\alpha}|$ from the greedy and Bayesian approaches. All $\hat{\alpha}$ are negative, indicating if two people have the same last name or social group membership, they are more likely to be linked. (In general, we plot the magnitudes to allow for easier comparison, particularly when we have both positive and negative $\hat{\alpha}_h$.) The larger the absolute value of $|\hat{\alpha}_h|$, the stronger the covariate effect is on the penalty. Here we see that the Bayesian approach gives more similar $\hat{\alpha}$ values across the ten runs, correctly identifying last name as the most important covariate and group B as having a slightly stronger effect than the other two groups. The greedy method gives $\hat{\alpha}$ values that are more varied and do not reflect the network design.  For example, although last name has a non-zero effect, it is not substantially larger than the other $\hat{\alpha}_h$ for the social groups.  One potential reason for the consistency differences is that the Bayesian approach is trying to optimize the log likelihood of $\alpha$ while the greedy algorithm tries to optimize the model performance based on the MSE which may find multiple combinations of $|\hat{\alpha}_h|$ that lead to similar results. For example, if two people share the same last name and the same social group, a smaller penalty on either last name or social group or both can help with recovering the link. 

%There are multiple combinations of $\alpha$ can lead to a relatively good performance on the overall network even though the $\alpha$ may not be the true $\alpha$. 

\begin{figure}[!ht]
    \centering
    \includegraphics[width=1\textwidth]{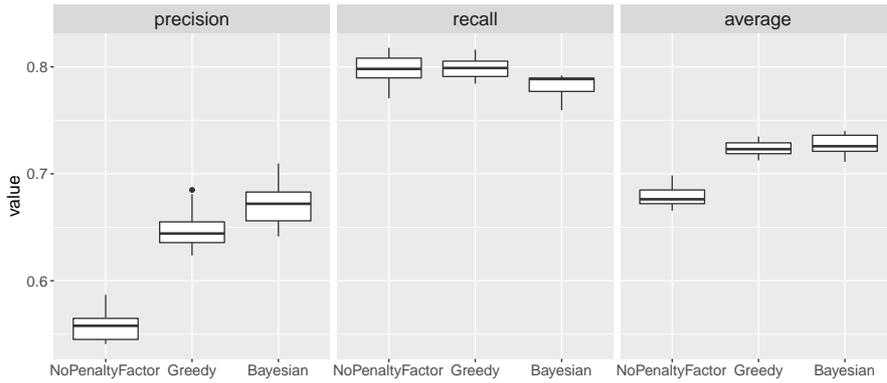}
    \caption{The distribution of precision and recall for the model without penalty factor, with penalty factor estimated using the greedy approach, and with the penalty factor estimated using the Bayesian approach. With penalty adjustment, both estimation approaches show an improvement in precision without a substantial change in recall.  There is no significant difference on the average of precision and recall between the two estimation approaches. }
    \label{fig:simu_result}
\end{figure}

For each generated document by person matrix, we also use the $\hat{\alpha}_h$ for both the greedy and Bayesian approaches to estimate the network and calculate the corresponding precision and recall, comparing these values to those for the model without penalty adjustment.  Figure  \ref{fig:simu_result} shows the resulting distributions for precision, recall, and the average of the two. We see that the model with penalty adjustment has improved precision, regardless of estimation approach, while the recall for all three options remains similar. 
The slight improvements in the average of the two follow.

%the estimated $\alpha$ for each data,  estimated the network and then calculate the precision and recall when the model without penalty adjustment, with the $\alpha$ estimated from greedy approach and the $\alpha$ estimated from Bayesian approach. 

%It seems that with a penalty adjustment, no matter the $\alpha$ is estimated from which approaches, the sum of precision and recall are largely improved. To be more specific, the precision is improved significantly without losing on recall. Moreover, there is no significant difference on the result of precision and recall with $\alpha$ estimated from different approach.

%Therefore, if we only want to improve the quality of predicted network, it is equivalent to use either approaches to estimate the $\alpha$. 

Now examining the predicted network structure, we see that all three model/estimation approaches overestimated the true number of links in the original simulated network (1164).  On average across the ten document by person matrices, the model without penalty adjustment detects 1662.1 links.  With penalty adjustment, the greedy estimation approach averages 1434.7 links, and the Bayesian approach averages 1358.1 links, both an improvement over the original model.

We then take a closer look at the estimated network structure for our ten family, 100 people sub-community (Figure \ref{fig:simu_network}) for two of the simulated document by person matrices. For Run 10 (top row of Figure \ref{fig:alpha}), the greedy and Bayesian estimation approaches give similar $|\hat{\alpha}_h|$ for last name, but the greedy approach gives slightly larger $|\hat{\alpha}_h|$ values (in magnitude) for the social group covariates. Therefore, we expect more links between people with the same group membership when using the estimates from the greedy approach compared to those of the Bayesian approach.  

\begin{figure}[!ht]
    \centering
    \includegraphics[width=1\textwidth]{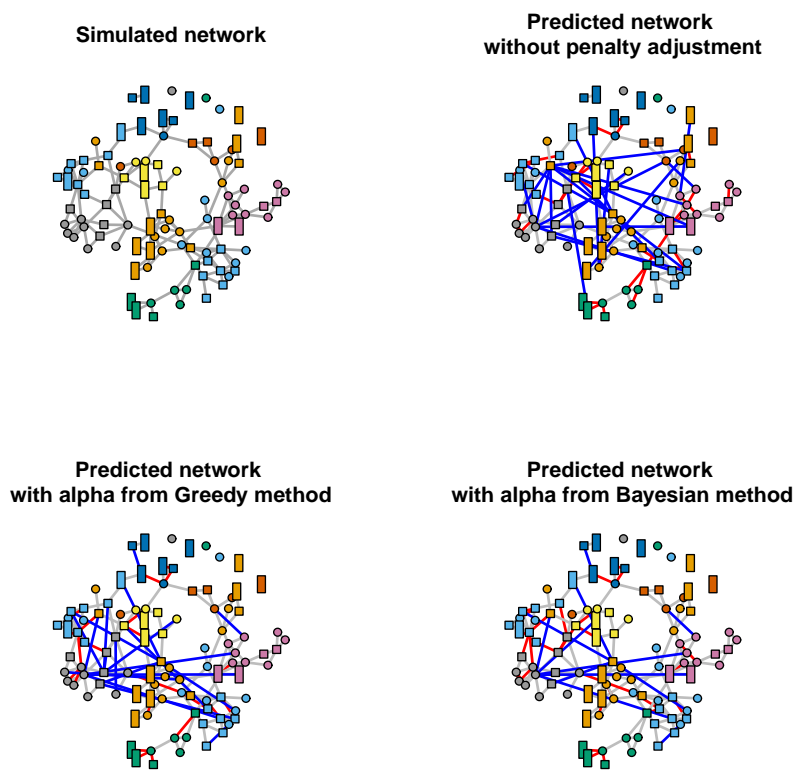}
    \caption{Predicted ten-family community network for Run 10 for the model without penalty factor, with the penalty factor estimated using the greedy approach and with the penalty factor estimated using the Bayesian approach. Grey line: True Positive; Blue line: False Positive; Red line: False negative. Node color: last name; Node shape: social groups. In Run 10, the $\hat{\alpha}_h$ are similar for last name; the greedy approach gives slightly higher values for group covariates.  We see fewer false positives using the greedy approach.}
    \label{fig:linkset1}
\end{figure}

The relevant estimated networks for Run 10 can be seen in Figure \ref{fig:linkset1}.  We can see that for the network estimated by the model without a penalty adjustment, the false positive links exists across the whole network but with for the network estimated with penalty adjustment, the number of false positive links decreases. The predicted networks with $\hat{\alpha}$ from the greedy and Bayesian approach are similar, but there are more within-group false positive links for this sub-community with $\hat\alpha$ from the greedy approach. Note that this is in line with the observation that the group-related penalties, for Run 10, are smaller for the greedy approach (and thus the within-group links are penalized less).

%We first take a look at run 10, which greedy and Bayesian gives a more similar estimation on $\alpha$ of last name, while greedy tends to puts more weights $\alpha$ of on groups. Thus, people with the same group membership are more likely to be connected. 

%for two of our d from each approach and thepredicted networks are plotted in Figure  and Fig. \ref{fig:linkset2}, respectively. 

\begin{figure}[!ht]
    \centering
    \includegraphics[width=1\textwidth]{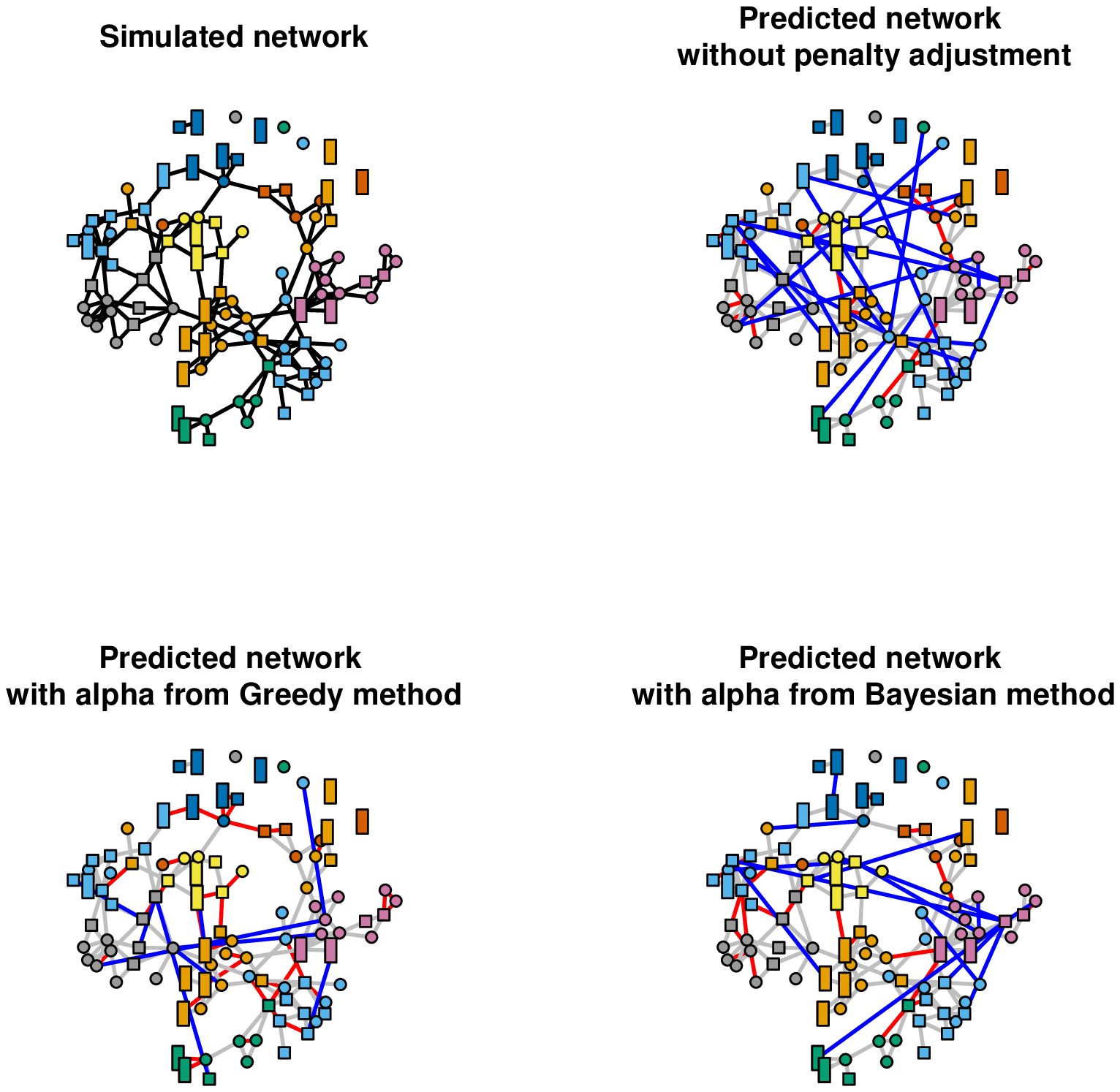}
    \caption{Predicted ten-family community network for Run 1 for the model without penalty factor, with the penalty factor estimated using the greedy approach and with the penalty factor estimated using the Bayesian approach. Grey line: True Positive; Blue line: False Positive; Red line: False negative. Node color: last name; Node shape: group membership. In Run 1, the greedy approach tends to give a smaller penalty on last name but a larger penalty on social groups which gives us more false positives and a few fewer false negatives within social groups.}
    \label{fig:linkset2}
\end{figure}

We also examine Run 1 where the $|\hat{\alpha}|$ are quite different between the two estimation approaches (see Figure \ref{fig:alpha}). The greedy method tends to give a smaller penalty for last name but a larger penalty on social groups, although we do note that the $|\hat{\alpha}_h|$ for group B is incorrectly estimated to be smaller than those for groups A, C. The corresponding predicted networks are depicted in Figure \ref{fig:linkset2}. The networks corresponding to the penalties estimated by the greedy and Bayesian approach are more dissimilar for for Run 1 than for Run 10, like the penalties themselves. Compared to Run 1, the last name and group B penalties are larger for the greedy approach, leading to the detection of fewer links between people with the same last name or who are both in group B.

%Notice that $\alpha$ estimated from Bayesian is an global estimation, thus the size of network is large, it will be slower than the greedy algorithm since it involves a matrix inverse calculation in Formula \ref{objective}. On the other hand, if the network is relatively smaller but the number of covariates is large, Bayesian is still preferred.  NEED TO INCORPORATE THIS IN SUMMARY

%and Bayesian approach tend to give a more accurate and consistent estimation on the penalty factor, which also corresponding to how the covariates are affecting the social network linkages. 

In summary, our simulation study gives some evidence that including covariate information through penalty adjustment can improve the performance of Local Poisson Graphical Lasso model in the context of estimating social networks from co-mention/count data derived from text. With respect to differences in the two estimation approaches, we see that the Bayesian approach tends to give more consistent results; however, we note that, given its global estimation and computational tasks (e.g. matrix inverse calculations), it will be slower than the greedy algorithm.

\section{Six Degrees of Francis Bacon: 1500-1575} \label{s:example}

We illustrate the model proposed in this paper by an application to part of the data used in the SDFB project \citep{warren2016six}, focusing on the period between 1500 and 1575. We compare the results of the models with and without covariate-dependent penalty factors. We consider the interpretability of the penalty factors, how they affect which network links are estimated, and approximate the precision of the models with and without penalty factors using Wikipedia as a reference. 

We first extract all documents from the SDFB database that contain references to individuals who were born and passed away between 1500 and 1575. This results in 2003 documents on 420 people. Over 83\% of them (394) are male, about 8\% (34) are female, and for the rest the gender is unknown. Women who appear in these data are usually associated with men in the data through family or marriage.

Apart from last name and birth and death year, we here consider three other covariates, related to individuals' occupation. We distinguish three groups: the Writer group (the occupation variable in the data contains the words ``poet", ``writer" or ``author"), the Church group (occupation contains ``church", ``religious", ``bishop" or ``catholic"), and the Royal group (occupation contains ``royal", ``king", ``queen" or ``regent"). 

Table \ref{tab} includes some descriptives of the data. Since we have limited the data to people who were alive in a period of 75 years, the lifespan of most pairs of people overlapped. Compared to the simulated data, the proportion of pairs with shared last name is much smaller. This indicates a more diverse last name distribution (the most common last name ``Stewart" is the last name of royalty during this period and appeared for only 9 individuals, while other last names appeared for no more than 5 individuals), but also suggests that as long as two people shared the same last name, the chances of them belonging to the same family and knowing each other are high. Among all occupations that were listed in the data, the writer and the church-related occupations are most popular. Individuals with a royal-related occupation tend to be closer connected than other people, which is why we consider this group, even though not that many people are part of it.
People can have multiple group membership across the three groups. Five individuals are part of more than one group, like Roger Ascham, who was an author and a royalty tutor, and John Seton, who was a Roman Catholic priest as well as a writer on logic.

\begin{comment}
\begin{table}[!h]
\caption{Descriptive statistics of the SDFB data on individuals from the period 1500--1575.}
\label{tab}
\begin{tabular}{ccc}
\hline\noalign{\smallskip}
Number of people  & Number of documents  & Gender ratio \\
\noalign{\smallskip}\hline\noalign{\smallskip}
 420 & 2003 &  349 (M):34 (F):37(unknown) \\
\noalign{\smallskip}\hline
\noalign{\smallskip}\hline\noalign{\smallskip}
Size of Poet/Writer group  & Size of Church group  & Size of Royal group \\
\noalign{\smallskip}\hline\noalign{\smallskip}
 40 & 49 & 19 \\
\noalign{\smallskip}\hline
\noalign{\smallskip}\hline\noalign{\smallskip}
People with multiple groups & \begin{tabular}[c]{@{}l@{}}Pairs of people who share \\the same last name \end{tabular} & \begin{tabular}[c]{@{}l@{}}Pairs of people with\\  birth/death year overlap \end{tabular}  \\ 
\noalign{\smallskip}\hline\noalign{\smallskip}
5 &  117 (0.133\%) & 81625 (92.77\%) \\
\noalign{\smallskip}\hline
\end{tabular}
\end{table}
\end{comment}

\begin{center}
\begin{table}
\caption{Descriptive statistics of the SDFB data on people from the period 1500--1575.}
\label{tab}
\begin{tabular}{lc}
\toprule
Group & Number of people \\
\midrule
\,\, Writer & 40 \\
\,\, Church & 49 \\
\,\, Royal & 19 \\
In multiple groups & 5\\
\midrule
  & Number of pairs \\ 
\midrule
\,\, Same last name & 117 \quad\,\,(0.13\%) \\
\,\, Lifespan overlaps & 81625 \,\,(92.8\%)\\
\bottomrule
\end{tabular}
\end{table}
\end{center}

We estimated the penalty parameters $\alpha$ using the Bayesian approach outlined in Section \ref{s:bayes}. We find that
\begin{equation}
    \begin{aligned}
    \alpha_{lastname} &= -1.853 \\
    \alpha_{writer} &= \phantom{+}0.369 \\
    \alpha_{church} &= -1.262 \\
    \alpha_{royal} &= -0.801.
    \end{aligned}
\end{equation}
From the size of the penalty factors, the last name is the most important covariate, indicating that if two people share the same last name, this is a strong indication that they may know each other. It is interesting that not for all groups the penalty factor is negative: if two people are both a writer, they are less likely to be connected. It is possible that being a writer is an occupation for which little collaboration is required, so that the writers did not socialize much with their peers. On the other hand, if two people are both related to the church or the royal family, this increases their chance of being linked. 

Next, we compare the networks generated by the Local Poisson Graphical Lasso model with and without penalty adjustment. The overall penalty level for both models is the one minimizing the MSE. For the model without penalty adjustment, the estimated network consists of 122 links and for the model with penalty adjustment, the estimated network consists of 99 links. Although they partially overlap, the two networks have also contain many different links. There are 36 links that are only picked up by the model with penalty adjustment picks up and 59 links that are picked up only by the model without penalty adjustment. 

How do the penalty factor values $\alpha$ relate to the difference between the two estimated network? To answer this question, we consider the percentage of links estimated by the two that had covariates in common (see Table \ref{tab_result}). 

\begin{table}[b]
\caption{Numbers and percentages of links estimated by the models with and without penalty adjustment, for whom the corresponding people had a common covariate.}
\label{tab_result}
\begin{tabular}{lcc}
\toprule
 & With penalty  &  Without penalty  \\
Covariate & adjustment model & adjustment model \\
\midrule
 Last name & 11 (11.1\%)  & 7 (5.7\%) \\
Writer group  & 1 (1.0\%)  & 1 (0.8\%) \\
Church group & 3 (3.0\%) & 4 (3.3\%) \\
Royal group & 7 (7.1\%) & 5 (4.1\%) \\ 
\bottomrule
\end{tabular}
\end{table}
As expected based on the negative penalty factor estimate for the last name and the royal group, the model with penalty adjustment picks up more links between people with the same last name or both related to the royal family. To be more specific, the model with penalty adjustment detects four additional links without losing the seven links that were estimated by the model without penalty adjustment model. However, the proportion of links between individuals from the writer or the church group does not differ much between the two models. Both models select one link between two people in the writer group. The model with penalty adjustment even picks one link less within the church group, even though the negative penalty factor $\alpha_{church}$ indicates that links between people within the church group are penalized less.  Note that  the difference in within-group estimated links only contributes a small portion of the difference among the estimated networks. This suggests that changing the penalty on the links between people within the same groups also affects the links that are not within those groups. 

Finally, we approximate the precision of the estimated networks by looking for evidence for links on Wikipedia. For a link involved with two people, as long one of the persons' Wikipedia document contains the other one's name, we consider this as evidence that a link exists. 
Of the 99 links that are picked by model with penalty adjustment, we find evidence for 59 (59.6\%). 
Of the 122 links that are picked by the model without penalty adjustment, we find evidence for 64 (52.5\%). 

Of the 59 links that are only picked by the model without penalty adjustment, there are 34 related to three people: Nicholas Bourbon, a French poet, and Margaret Roper and Katherine Seymour, two noblewomen. Nicholas Bourbon only spent one year in the United Kingdom during his life. Even though he did have a strong connection to Anne Boleyn, the Queen Consort, there is no other evidence to suggest that he is connected to the Queen's other acquaintances, even though he may have mentioned those people in his series of poems about the Queen. We find a similar pattern for the two noblewomen: in reality, they did have some strong connections to royalty and other aristocrats, but the model without penalty adjustment incorrectly links them to people who they are connected to only indirectly. The model with penalty adjustment did not detect the links between the noblewomen and the royal group either. Modifying the definition of the royal group or adding a group for nobility could help improve the model's performance. 

There is no doubt that an in-depth analysis of these results would require the help from experts on British history, but from these preliminary analyses, it seems that the model with penalty adjustment yields a more precise and conservative estimate of the relationships.

\section{Discussion}

In this paper, we have shown promising results to support adding covariate information when estimating social networks from text data using a Local Poisson Graphical Lasso model. This covariate information is incorporated through the L1 penalty: we penalize the parameters representing the edges between two individuals depending on the extent to which they have covariates in common. To estimate the penalty factors, we have discussed two approaches:  a greedy algorithm and a Bayesian framework.

Our simulation results indicate that, if there is relevant covariate information, including this in the L1 penalty can improve the sum of precision and recall of the estimated network. We did not observe any significant differences in precision or recall when the penalty factor was estimated by either the greedy or the Bayesian approach.  While the greedy approach typically requires less computation time, it does require a prior decision about a reasonable search range for the penalty factor.  The Bayesian approach tends to give a more consistent estimate for the penalty factors, but requires more computation time. In the Bayesian approach, we do approximate both the Poisson regression likelihood and the Laplace prior to find the marginal likelihood for the penalty factor. We have not analytically evaluated the effect of this double approximation. However, the simulation study yields that the penalty factor estimated by the Bayesian approach gives results comparable to those for the greedy approach.

We applied the Local Poisson Graphical Lasso model with penalty adjustment to biographical data on individuals who lived in early modern Britain in the period 1500 to 1575, estimating the penalty factors using the Bayesian approach. We find that in these real data, the relation between covariates and linkage probability is not always in line with homophily theory. While individuals with the same last name or with both a church- or royal-related occupation were more likely to be linked, according to the estimated penalty factors, links between individuals who we both authors were actually penalized more heavily. Also, based on Wikipedia as an external reference, we conclude that the model with penalty adjustment could help to recover the social network with higher precision compared to the model without penalty adjustment.
%(that is, if the two models would achieve a similar level of recall, like in the simulation). 
Here, we have applied the model to a subset of the SDFB data. The complete SDFB data contain over 19000 documents with references to over 13000 people. Both approaches we have proposed to estimate the penalty factor, and especially the Bayesian approach, will be slow when dealing with large data. Considering other optimization approaches to improve the computational efficiency is an interesting avenue for future research. 

We only interpret the positive edge coefficients in the Local Poisson Graphical Lasso model as an indication for the existence of a social tie between to individuals, %thus the coefficients should be limited to be non-negative and that is why we force 
and have assumed the coefficients to be non-negative in the estimation. When we ran the same models without this constraint and only consider the positive coefficients as an indication of a link, we obtained roughly the same results. However, alternatively, we could have opted for a different model or a different penalty structure/prior, that would do away with the post-processing step.  For example, it could be interesting to explore the Bayesian approach with a combination of a Poisson regression likelihood and a Gamma prior.

Finally, in this paper, we only focus on binary covariates. However, there are several continuous covariates that are worth considering when estimating social networks from text data. For example, historical text often contains  typos and sometimes spelling variation. Therefore,  when identifying whether two people are from the same family, instead of directly comparing their last names, we could consider the similarity of their last names. By measuring name similarity using the Jaro-Winkler distance \citep{winkler1990string} and including this in the L1 penalty, a missing letter or a word with the same pronunciation but a different spelling (e.g., ``Askham" and ``Ascham") will not be over-penalized. Also, in the real data example, we currently only consider how being in the same occupation group should affect the L1 penalty. However, this analysis does not need to be limited to the within- versus between-occupation comparison. We could also include a continuous covariate representing the similarity between occupations  (or introduce penalty parameters corresponding to the links between occupation groups). The proposed Local Poisson  Graphical  Lasso  model with covariate-dependent penalty parameters thus provides a rich framework for learning social networks from text data.

\newpage
\section*{Appendix A: Approximating the marginal likelihood of the penalty parameters}

As mentioned in equation \eqref{eq:model}, we model the number of times person $j$ appears in document $i$ using Poisson regression,
\begin{equation}
  Y_{ij}\,|\,Y_{i,\neq j} = y_{i,\neq j}, \theta, \Theta \sim \text{Poisson}(e^{\lambda(y_{i, \neq j})}),
  \end{equation}
where
\begin{equation}
\lambda(y_{i, \neq j}) = \theta_j + \sum_{k \neq j} y_{ik} \Theta_{kj}, 
 \end{equation}
with a covariate-depedent Lasso penalty on the $\Theta_{kj}$ or, equivalently, a Laplace prior (see expression \eqref{eq:laplaceprior}).
To estimate the values of penalty parameters $\alpha$ in a Bayesian framework, we here approximate their marginal likelihood. 

We first approximate the Poisson likelihood by a normal distribution, using the log-gamma approximation \citep{bartlett1946statistical,prentice1974log,chan2009bayesian}. Recall a Gamma random variable $\mu \sim \text{Gamma}(a, b)$ with distribution
\begin{equation}
    p(\mu\mid a,b) = \frac{1}{\Gamma(a)b^a} \mu^{a-1}\exp^{-\frac{\mu}{b}},
\end{equation}
then the transformed random variable $\log(\mu)$ has a log-gamma distribution and for large $a$,  the log-gamma distribution is approximately to $\mathcal{N}(\log(a)+\log(b), a^{-1})$. Let $b=1$ and $a \in \mathcal{Z}^+$, and let $\eta = \log(\mu)$ then we have
\begin{equation}\label{eq:approx}
\begin{aligned}
    p(\eta\mid a,1) & = p(\mu = \exp^\eta\mid a,1)\times \frac{\partial}{\partial \eta} \exp^\eta \\
    & = \frac{1}{(a-1)!} \exp^{\eta a} \exp^{-\exp^\eta} \\
    & \approx \text{G}(\eta \mid \log(a), a^{-1}),
    \end{aligned}
\end{equation}
where $G(x|\mu, \Sigma) = (2\pi)^{-d/2} |\Sigma|^{-1/2} \exp(\frac{1}{2}|| x - \mu||^2_\Sigma)$
is the equation of a multivariate Gaussian distribution, $||x||^2_\Sigma = x^T \Sigma^{-1} x$.

Since our data $y$ are often sparse, to avoid $\log(0)$ in the remainder of this derivation, we add 1 to all response values $y_{ij}$ and define $y_{ij}^* = y_{ij} +1$.
Using the approximation derived in equation \eqref{eq:approx}, we find that 
%When applying such transformation to our application, since our data matrix $y$ are relatively sparse, so in order to avoid the $\log(0)$, we add 1 to all response values as 
\begin{equation}
   % \begin{aligned}
%    p(\lambda(y_{i, \neq j}) |y_{ij},1) &= p(\mu = \exp^{\lambda(y_{i, \neq j})}|y_{ij},1)\frac{\partial}{\partial\lambda(y_{i, \neq j}) } \exp^{\lambda(y_{i, \neq j}) } \\
     \frac{1}{(y_{ij}^*-1)!} e^{\lambda(y_{i, \neq j}) y_{ij}^*} e^{-e^{\lambda(y_{i, \neq j})}} \\ 
    %&\approx \frac{1}{y_{ij}!} \exp^{\lambda(y_{i, \neq j}) (y_{ij}+1)} \exp^{-\exp^{\lambda(y_{i, \neq j})}} \\
    \approx \text{G}\left(\lambda(y_{i, \neq j}) \mid \log(y_{ij}^*), \frac{1}{y_{ij}^*}\right). 
 %   \end{aligned}
\end{equation}
%where $\Sigma_j = diag[\frac{1}{y_{1j}}, ... \frac{1}{y_{nj}}]$. 
Then the Poisson likelihood can be written as
\begin{equation}
    \begin{aligned}
    \prod_{i=1}^n p(Y_{ij}^*\mid Y_{i, \neq j} = y_{i, \neq j}, \Theta_{ j}) &= \prod_{i=1}^n \frac{1}{y_{ij}^* !}e^{\lambda(y_{i, \neq j}) y_{ij}^*} e^{-e^{\lambda(y_{i, \neq j}) }} \\
    &\approx \prod_{i=1}^n \text{G}\left(\lambda(y_{i, \neq j}) \mid \log(y_{ij}^*), \frac{1}{y_{ij}^*}\right) \\
    %&\approx \frac{|\sigma_j I_n| ^{-1/2}}{(2\pi)^{\frac{N}{2}}} e^{-\frac{1}{2}|| \lambda(y_{ \neq j})- \log(y_{j} + 1)||^2_{\sigma_j I_n}}
    \end{aligned}
\end{equation}
Since the variance $1/y_{ij}^*$ is likely to be similar across the documents $i$, we set  $\hat\sigma^2_j = \frac{1}{n} \sum_{i=1}^n \frac{1}{y_{ij}^*}$.

At this point, we specify the Laplace prior defined in equation \eqref{eq:laplaceprior} by 
\begin{equation}
    \Theta_{kj} \sim \text{Laplace} \left(0,\frac{\rho_{kj}}{2\sigma_j^2}\right).
\end{equation}
We approximate this Laplace prior by a normal distribution with the same variance \citep{zeng2020incorporating}. That is, we can approximate $\beta_j \sim \mathcal{N}(0, \frac{2}{\tau_j^2})$ by $\beta_j \sim \text{Laplace}(0, \tau_j)$. Therefore, we can approximate the distribution of edge parameters $\Theta_{\neq j, j}$ by
\begin{equation}
\Theta_{\neq j, j} \sim \mathcal{N}(0, V^j)
\end{equation}
where $V^j \in \mathbb{R}^{(p-1)\times (p-1)}$ is a diagonal matrix with $V^j_{kk} = \frac{2\sigma_j^2}{e^{2Z^{j*}_k\alpha}} $ in which $Z^{j*}_k$ is the $k$th row of the covariate matrix $Z^{j*}$. Now we can write out the marginal likelihood of $\alpha$:
\begin{equation}
\begin{aligned}
L_j(\alpha) &= \int_{\mathbb{R}^p} \prod_{i=1}^n p(Y_{ij}^*\mid Y_{i, \neq j} = y_{i, \neq j}, \Theta_{j}) \prod_{k\neq j} p(\Theta_{kj}\mid\alpha) \dd \Theta_{j} \\
&= \int_{\mathbb{R}^p} \prod_{i=1}^n \frac{1}{y_{ij}^* !}e^{\lambda(y_{i, \neq j}) y_{ij}^*} e^{-e^{\lambda(y_{i, \neq j}) }}  \prod_{k\neq j} \frac{e^{(Z^{j*}_k\alpha)}}{4\sigma_j^2} e^{-\frac{e(Z^{j*}_k \alpha)}{2\sigma^2}|\Theta_{kj}|} \dd \Theta_{j} \\
&\approx \int_{\mathbb{R}^p} \frac{|\sigma_j I_n| ^{-1/2}}{(2\pi)^{\frac{N}{2}}} e^{-\frac{1}{2}|| \lambda(y_{ \neq j}) - \log(y_{j}^*)||^2_{\sigma_j I_n}} \frac{|V^j|^{-1/2}}{(2\pi)^{\frac{p}{2}}} e^{-\frac{1}{2} ||\Theta_j||_{V^j}} \dd \Theta_{j}.
\end{aligned}
\label{alphalikelihood}
\end{equation}
where $y_{j}^* = (y_{1j}^*, \ldots, y_{nj}^*)^\top$, $y_{\neq j}$ denotes $y$ excluding the $j$th column, and within the norm, $\lambda(\cdot)$ operates on the columns of $y_{\neq j}$ and $\log(\cdot)$ is applied element-wise. Dropping terms that are not a function of $\Theta_{j}$, expanding the norm term and completing the square within the integral, we obtain that the approximate marginal log-likelihood of $\alpha$ satisfies
\begin{equation}
    -l_j(\alpha) \propto \log|C_{\alpha}| + \log(y_{j}^*)^\top C_{\alpha}^{-1} \log(y_{j}^*)
\end{equation}
where $C_\alpha = \sigma_j I^2 + y_{\neq j} V^{j}y_{\neq j}^{\top}$.

%\begin{acknowledgements}
%\end{acknowledgements}

\bibliographystyle{spbasic}      % basic style, author-year citations
\bibliography{main}   % name your BibTeX data base

%\begin{thebibliography}{}
% etc
%\end{thebibliography}

\end{document}